\documentclass[prl,aps,twocolumn,groupedaddress,showpacs,floatfix]{revtex4}

\usepackage{amsmath,amssymb,multirow,epsfig,bm}

\newcommand{\etal}{{\it et al.,\;}}
\newcommand{\beq}{\begin{equation}}
\newcommand{\eeq}{\end{equation}}
\newcommand{\bea}{\begin{eqnarray}}
\newcommand{\eea}{\end{eqnarray}}

\newcommand{\veps}{\varepsilon}

\newcommand{\nn}{\nonumber}
\newcommand{\benn}{\begin{displaymath}}
\newcommand{\eenn}{\end{displaymath}}
\begin{document}

\title{\bf On the specific heat of a fermionic atomic cloud in
the unitary regime }

\author{ Aurel Bulgac}
\affiliation{Department of Physics, University of
Washington, Seattle, WA 98195--1560, USA}

\date{\today}

\begin{abstract}

In the unitary regime, when the scattering amplitude greatly exceeds
in magnitude the average inter-particle separation, and below the
critical temperature thermal properties of an atomic fermionic cloud
are governed by the collective modes, specifically the
Bogoliubov-Anderson sound modes. 
The specific heat of an atomic cloud in a elongated 
trap in particular has a rather compex temperature dependence, which
changes from an exponential behavior at very low temperatures
($T\ll\hbar\omega_{||}$), to $\propto T$
for $\hbar\omega_{||}\ll T \ll \hbar\omega_\perp$
and then continuosly to $\propto T^4$ at temperatures
just below the critical temperature, when the surface modes play a dominant role.
Only the low ($\hbar\omega_{||} \ll  T \ll \hbar\omega_\perp$) and high 
($\hbar\omega_\perp \ll T < T_c$) temperature power laws are well defined. For the intermediate
temperatures one can introduce at most a gradually increasing with temperature
exponent.

\end{abstract}


% 03.75.Ss Degenerate Fermi gases
 .
\maketitle

%----------------------------------------------------------
%----------------------------------------------------------

Dilute atomic Fermi gases \cite{exp0}, and especially dilute gases
interacting with large scattering lengths are under intense scrutiny
both experimentally \cite{exp1,exp2,thomas} and theoretically
\cite{theory,carlson,chang,giorgini,amoruso,oscill,stringari,abgfb05,ho,heiselberg,levin}
(these lists, especially the list of theoretical contributions, is far
from being exhaustive or, maybe, even representative).  While quite a number
of properties of these systems have been clarified, the full and exact
picture still awaits to be uncovered. What was not appreciated so far
is the fact that from the information inferred so far, both
theoretically and experimentally, one can draw a number of unambiguous
conclusions concerning the thermal properties of these systems. It is
widely accepted that the equation of state at zero temperature is
known theoretically with an accuracy of a few percent
\cite{carlson,chang,giorgini}. This equation of state allows us also
to derive the spectrum of low lying collective excitations of such systems
\cite{amoruso,oscill,stringari,abgfb05}, namely the sound waves, and
one can safely assume that experimentally \cite{exp2} the basic
properties of these collective modes agree with theory. I shall show
here how this information can be profitably used in order to extract
information about the thermal properties of these systems at
temperatures below the critical temperature for transition into a
normal state. I shall also show how the geometry of the these clouds
determines in a somewhat unexpected manner the specific heat of these
clouds. In the first part of this work I shall discuss briefly some of the
properties of the infinite homogeneous systems and I shall turn to the
discussion of clouds in harmonic traps in the second part. Various
aspects of the thermal properties of fermionic clouds have been
considered by others before \cite{ho,heiselberg,levin}, however, as
far as I am aware of, the role of the collective excitations below the
pairing gap has not been discussed previously in this context.

In an infinite Fermi system with number density $n=k_F^3/3\pi^2$, in
the unitary regime, the speed of the Bogoliubov-Anderson sound mode is
given by \cite{abgfb05}
%----------------------------------------------------------
\beq \label{eq:c2}
c^2 = \frac{\xi_s v_F^2}{3}
\left [ 1 -\frac{2\zeta_s}{5\xi_s k_Fa} +
{\cal{O}}\left(\frac{1}{(k_Fa)^2}\right )\right ],
\eeq
%----------------------------------------------------------
where $v_F=\hbar k_F/m$ and $k_F$ are the Fermi velocity and wave vector
respectively and $a$ is the scattering length. The Bogoliubov-Anderson
sound waves have the expected linear dispersion law with momentum
$\hbar k$, namely $\omega_s = ck$.  The dimensionless parameters
$\xi_s\approx 0.44$ and $\zeta_s\approx 1$ have been determined recently
\cite{carlson,chang,giorgini}. At low temperatures one can expect 
that only two types of elementary excitations exist, the boson-like 
Bogoliubov-Anderson phonons and the fermion-like gapped Bogoliubov quasi-particles.
One can estimate their contribution to the total energy $E(T)$ by
assuming that at $T=0$ the system is a Fermi superfluid with a 
ground state energy and a pairing gap determined in Ref. \cite{carlson}.
One thus obtains \cite{bdm}, see Ref. \cite{fw} for derivations of such formulas:
%----------------------------------------------------------
\bea 
E_s(T) &=& \frac{3}{5}\varepsilon_F N\left [
\xi_s +  \frac{\sqrt{3}\pi^4}{16\xi_s^{3/2}}
\left ( \frac{T}{\varepsilon_F}\right ) ^4 \right . \nn \\
  &+& \left .\frac{5}{2}\sqrt{\frac{2\pi\Delta^3T}{\veps_F^4}}
\exp\left (-\frac{\Delta}{T}\right)
\right ], \label{eq:Es}
\eea
%----------------------------------------------------------
where $\varepsilon_F =\hbar^2k_F^2/2m$, $N$ is the total number of 
particles and the temperature is on the
energy scale $(k_B=1)$ and $\Delta$ is the pairing gap at $T=0$.
As illustrated in Fig. 1 of Ref. \cite{bdm} at $T\approx 0.2 \veps_F$ the 
two contributions are essentially equal (each of the last two terms 
inside the square brackets $\approx 0.06$).  
Off the resonance (when $1/k_Fa \ne 0$), one has to include corrections
similar to those in
Eq. (\ref{eq:c2}). Only the (mean-field) exponentially suppressed
contribution, due to the breaking of the Cooper pairs, and has
been considered previously \cite{heiselberg,levin}. Above the
critical temperature, such a system behaves as a normal Fermi system
and its energy can be estimated with the usual textbook formula
%----------------------------------------------------------
\beq\label{eq:En}
E_n(T) = \frac{3}{5}\varepsilon_F N\left [ \xi_n +
\frac{5\pi^2}{12}\left(\frac{T}{\varepsilon_F}\right)^2 \right ],
\eeq
%----------------------------------------------------------
where $\xi_n\approx 0.59$ can be estimated from the condensation
energy or be computed directly \cite{chang0,carlson}.  Using these two
expressions, Eqs. (\ref{eq:Es}) and (\ref{eq:En}), one can show that
the corresponding free energies cross at a temperature $T_{cross}
\approx 0.2\varepsilon_F$. This fact would suggest that the unknown so
far critical temperature of a such a system is around
$0.2\varepsilon_F$. This estimate is noticeably lower than what a
mean-field or BCS-like treatment would suggest in the weak
coupling limit, namely $T_{c} \approx 0.5\varepsilon_F$
\cite{theory,leggett}. Notice that even in the weak coupling limit the
BCS theory overestimates of the critical temperature \cite{gorkov},
and the magnitude of this correction seems to be the same in the
strong coupling limit \cite{chang}.  The above estimate for the 
critical temperature is rather close to a recent
experimental claim \cite{thomas}. 
The theoretical mean-field analysis performed in conjunction
with this experiment \cite{levin}, did not consider the role
played by the collective modes in the thermal properties for $T <
T_c$.  As it will be shown below, in a trap
the thermal properties of a fermion cloud  are modified in a qualitative
manner and are expected to show the presence of several
different regimes.

In typical experiments with cold fermionic atoms the trap is largely
harmonic, but it also has a very elongated shape. Since
$\hbar\omega_{||} \ll \hbar\omega_\perp \ll \varepsilon_F$, the shape
of the atomic cloud is that of a very long cigar. The spectrum of the
collective oscillations of such a superfluid systems in a spherical
trap have been evaluated in Refs. \cite{amoruso}
%-------------------------------------------------------------
\beq \label{eq:osph}
\Omega_{nl} = \omega \sqrt{ \frac{4}{3}n(n+l+2) +l },
\eeq
%-------------------------------------------------------------
where $\omega$ is the frequency of the trap and $n$ and $l$ are the
radial quantum number and $l$ is the corresponding angular
momentum. One notices that even though the spectrum is distinctly
different from a spherical harmonic spectrum, nevertheless it is quite
close to one (up to the notorious scale factor $1/\sqrt{3}$), namely
%-------------------------------------------------------------
\beq \label{eq:oappr}
\Omega_{nl} \approx \frac{\omega}{\sqrt{3}}\left ( 2n+l+\frac{3}{2}\right ).
\eeq
%-------------------------------------------------------------
This behaviour is similar to the change in dispersion from 
$\omega = v_Fk$ for a homogeneous noninteracting Fermi gas to 
$\omega = v_Fk/\sqrt{3}$ for a homogeneous Fermi superfluid.
Since I shall be interested in global properties of these systems,
which depend on the propperties of the entire spectrum of these oscillations, the
differences between the exact and approximate spectra, naively,
are not expected to change the qualitative picture. For
temperatures $T \ll T_c ={\cal{O}}(\Delta)={\cal{O}}(\varepsilon_F)$
only such collective oscillations can be excited with a noticeable
probability. The energy of an
atomic fermionic elongated cloud is then clearly given by the formula
%-------------------------------------------------------------
\bea \label{eq:etrap}
& &  \!\!\!\!\! E_s(T) = E(0) + \sum _{n_x,n_y,n_z}^\prime
\frac{\hbar\Omega(n_x,n_y,n_z)}{\exp[\beta\hbar\Omega(n_x,n_y,n_z)] -1},\\
& &\Omega(n_x,n_y,n_z)= \frac{1}{\sqrt{3}}
\left [\omega_{\perp}(n_x+n_y) +n_z\omega_{||} \right ],
\eea
%-------------------------------------------------------------
where $\beta = 1/T$ and the prime superscript is a reminder that only
terms with $n_x+n_y+n_z>0$ are to be included in the sum. Since the
longitudinal $\omega_{||}$ and radial $\omega_\perp$ frequencies
differ typically by a factor close to a hundred, one can distinguish several
temperature regimes.  For very low temperatures, when $T\ll
\hbar\omega_{||}$, the sum in the above formula is exponentially small
in comparison with $\hbar\omega_{||}$ and the cloud is essentially in
the ground state. The second regime is for temperatures in the
interval $\hbar\omega_{||} \ll T \ll\hbar \omega_\perp$, when one can
easily show that
%-------------------------------------------------------------
\beq \label{eq:e2}
E_s(T)\approx E(0) + \frac{\sqrt{3}\pi^2}{6}\frac{T^2}{\hbar\omega_{||}}.
\eeq
%-------------------------------------------------------------
For temperatures in the interval $\hbar\omega_\perp\ll T \ll T_c$ one
can show that the total energy is essentially quartic in temperature,
namely
%-------------------------------------------------------------
\beq \label{eq:e4}
E_s(T)\approx E(0) + \alpha\frac{T^4}{\hbar^3\omega_{||}\omega_\perp^2},
\eeq
%-------------------------------------------------------------
where the constant $\alpha$ can be estimated as the integral
%-------------------------------------------------------------
\bea
&& \alpha = 3^{3/2}\int_0^\infty dx \int_0^\infty dy\int_0^\infty dz
\frac{x+y+z}{\exp(x+y+z)-1} \nn \\
& & = \frac{3^{3/2}\pi^4}{30} = 16.87.
\eea
%-------------------------------------------------------------
Above $T_c$ one expects a normal Fermi liquid behavior, with an essentially
$T^2$ dependence. Naturally, in this rough picture the detailed
critical behavior at and around the critical point is neglected. The
fact that the energy as a function of temperature can change its
character is known in anisotropic solids for a long time \cite{LL},
however it was not appreciated yet in the context of fermionic atomic
clouds in anisotropic traps so far. It is important to realize that
the geometric shape of the could alone can lead to this somewhat
unexpected behavior of $E(T)$, and of the specific heat as well, and
that no phase transition happens really.  The transition from the
regime corresponding to Eq. (\ref{eq:e2}) to the one corresponding to
Eq. (\ref{eq:e4}) is not abrupt, but occurs over a rather long
temperature interval, and can be easily described with a temperature
dependent exponent.

The approximate spectrum in Eq. (\ref{eq:oappr}) does not reproduce
correctly the dispersion of the surface modes, namely
$\Omega_{surf}\approx \omega\sqrt{(4n+3)l/3}$ (for $n\ll l$) 
and, as it will shown below, their role in the thermal properties of 
a cloud is somewhat unexpected.
One can estimate $E(T)$ by evaluating the sum over the quantum numbers
$(n,l,m)$. For temperatures $T\gg \Omega$ one then obtains
%-------------------------------------------------------------
\bea
& & E_s(T)= E(0) + \sum_{n,l}^\prime
\frac{(2l+1)\hbar\Omega_{nl}}{\exp(\beta\hbar\Omega_{nl})-1} \nn \\
& & \approx  \frac{\xi_s\hbar\omega (3N)^{4/3}}{4}+
142 \frac{T^5}{\hbar^4\omega^4}.
\eea
%-------------------------------------------------------------
Both the temperature exponent and the prefactor have been obtained
numerically. One can evaluate this expression also by replacing the
sum over $l$ with an integral, namely
%-------------------------------------------------------------
\bea
& & E_s(T)\approx E(0) + 
\sum_{n=0}^\infty 2\hbar\omega\sqrt{1+\frac{4n}{3}} \nn \\
& &\times \int dl l^{3/2}
\exp\left (-\frac{\hbar\omega l^{1/2}}{T}\sqrt{1+\frac{4n}{3}}  \right )
\nn \\
& & \approx
0.25\xi_s^{1/2}\hbar\omega (3N)^{4/3} + 140 \frac{T^5}{\hbar^4\omega^4}.
\label{eq:surf}
\eea
%-------------------------------------------------------------
Naturally, there are sub-leading corrections to these estimates,
notably corrections of order ${\cal{O}}(T^4)$. 
It is desirable to extend this formula to the case of an axially 
deformed trap. Unfortunately a simple analytical formula for the 
collective modes is not available in this case \cite{cigar}. 
One can use the fact that the (local) frequency $\Omega_{surf}(S)$
of the surface modes with wave vector $k$ 
can be determined rather accurately from the classical formula
%-------------------------------------------------------------
\beq
\Omega^2_{surf}(S) = k\frac{F(S)}{m},
\eeq
%-------------------------------------------------------------
where $F(S)=|{\bf \nabla} U({\bf r})|_S$ is the force acting on a particle at 
the surface of the cloud, see Refs. \cite{surf} for a detailed 
discussion. In the above formula the argument $S$ stands for a particular point
on the cloud surface.
Using an axially symmetric harmonic trapping potential 
$U({\bf r})=m\omega^2(x^2+y^2+\lambda^2z^2)/2$ and the 
semiclassical formula for the contribution of the 
surface modes to the total energy
%-------------------------------------------------------------
\bea
&&\!\!\!\!\!\!\!\!\!\!\!\!\!\!\!\!
 E_s(T)\approx E(0) + \int \frac{dS d^2k}{(2\pi)^2}
\frac{\hbar\Omega_{surf}(S)}{\exp[\beta\hbar\Omega_{surf}(S)]-1}\nn\\
&& \!\!\!\!\!\!\!\!\!\!\!\!\!\!\!\!
\approx 0.25\xi_s^{1/2}\hbar\omega (3N)^{4/3} + 
\frac{96 T^5}{\hbar^4\omega^4}
\quad \frac{\arctan{\sqrt{\lambda^2-1}}}{\lambda\sqrt{\lambda^2-1}}.
\eea
%-------------------------------------------------------------
Here $\int dS$ stands for the integral over the surface of the cloud.
In the case of a spherical trap ($\lambda=1$) one recovers 
the previous formula, see Eq. (\ref{eq:surf}), if one includes only the modes with 
$n=0$. One can thus conclude that the effect of the effect of the deformation of 
the trapping potential can be encapsulated in a simple shape factor. Since 
most experiments are performed in  cigar-like traps ($\lambda \ll 1$), the role of 
the surface modes is even larger ($\propto 1/\lambda$), when compared to
spherical traps.

It is useful to
obtain an estimate of the maximum temperature up to which this
formula is expected to be valid. One assumption is that
$T < \Delta(T)$, which is likely to hold up to temperatures
$T\approx T_c/2$. A reasonable estimate of the important angular
momenta in the above integral is
%-------------------------------------------------------------
\beq
l \approx \left ( \frac{T}{\hbar\omega} \right ) ^2 < l_{max}\approx
(24N)^{1/3},
\eeq
%-------------------------------------------------------------
where $l_{max}$ is a rough estimate of the largest single-particle angular
momentum of a fermion in the ground state. Using this estimate one
obtains for the total energy the following rough estimate
%-------------------------------------------------------------
\beq
E_s(T) <
\frac{\xi_s^{1/2}\hbar\omega (3N)^{4/3}}{4} +
140\hbar \omega (24N)^{5/6}.
\eeq
%-------------------------------------------------------------
Perhaps the thermal energy stored in the surface modes is grossly
overestimated. At high, but not yet determined, 
temperatures the surface will ``melt'' and the $T^5$ regime
will be replaced by a normal Fermi liquid behavior at some point.  While a good
case could be made that the exact numerical factor in front of the
second term in the above formula is too large, the dependence on the
particle number is probably correct. One can easily estimate the free
energy as well and obtain that
%-------------------------------------------------------------
\bea
F_s(T) &=&\frac{\xi_s^{1/2}\hbar\omega (3N)^{4/3}}{4}
-35 \frac{T^5}{\hbar^4\omega^4} \nn\\
&>&
\frac{\xi_s^{1/2}\hbar\omega (3N)^{4/3}}{4} -
35\hbar \omega (24N)^{5/6},
\eea
%-------------------------------------------------------------
which would have to be contrasted with the free energy of the normal phase
%-------------------------------------------------------------
\beq
F_n(T) =\frac{\xi_n^{1/2}\hbar\omega (3N)^{4/3}}{4} -
\frac{\pi^2(24N)^{2/3}T^2}{24\xi_n^{1/2}\hbar\omega}.
\eeq
%-------------------------------------------------------------
This estimate of the free energy of the normal phase implies that no
collective modes exist. Since the interaction between fermions is
attractive, the Landau's zero sound mode is unstable. There is
still the question of whether the first sound mode can exist in such a
system at relatively low temperatures in the normal phase. If the first
sound modes would be stable, then their contribution to the free
energy would be very similar to the contribution of the
Bogoliubov-Anderson sound modes to the free energy of the superfluid
phase. So far neither theoretical nor experimental evidence of a
first sound mode in such systems in the unitary regime exist.  This
fact should be not construed, however, as evidence of its
non-existence.

If no first sound modes exist, then the inescapable conclusion emerges
that the free energy of the superfluid phase is likely well below the
free energy of the normal phase in a trap for a temperature range much
larger than in the bulk. That would mean that an atomic fermionic
cloud in the unitary regime remains superfluid at temperatures larger
than the bulk critical temperatures.
In the above analysis I have obviously neglected the contribution due
to thermally broken Cooper pairs. This contribution, which 
has been considered previously by others, was not the subject of 
this work.

In conclusion, the thermal properties of fermionic atomic gases,
especially in the unitary regime, when the scattering length greatly
exceeds in magnitude the average inter-particle separation, are
dominated by the thermal excitation of the Bogoliubov-Anderson sound
modes, both in the bulk and in traps. In traps, the specific heat of
such a system, at temperatures below the critical transition to a
normal phase, has a rather complicated temperature dependence, ranging
from exponentially damped at very low $T$'s to various power laws.  At
temperatures higher than the axial frequency of an elongated trap the
specific heat has a linear in $T$ regime, due to the excitation of
predominantly axial sound modes. At higher temperatures, close but 
below the critical temperature, this regime is
replaced with a regime in which $C\propto T^4$, when the excitation of
pure surface modes dominate the thermal properties of atomic clouds in
traps. This behavior is obviously absent in the bulk and is
characteristic of finite systems only. A similar behavior of the
specific heat is also expected in the case of an atomic Bose gas in a
trap, since the spectrum of surface mode is similar
\cite{alan}.

A number of conversations with my colleagues G.F. Bertsch and
D.J. Thouless have been very helpful for clarifying some of my initial
ideas. I thank J.E. Drut for correcting a slip, P. Magierski for
reading the draft and D.T. Son for discussions.  This work was
supported in part by the Department of Energy under grant
DE-FG03-97ER41014.

%-------------------------------------------------------------
%-------------------------------------------------------------


\begin{thebibliography}{99}

\bibitem{exp0} B. DeMarco and D.S. Jin, Science, {\bf 285}, 1703
(1999).

\bibitem{exp1} K.M. O'Hara {\it et al.}, Phys. Rev. Lett. {\bf 82},
4204 (1999); K.M. O'Hara, {\it et al.,} Science, {\bf 298}, 2179
(2002); M.E. Gehm, {\it et al.,} Phys. Rev. A {\bf 68}, 011401 (2003);
T. Bourdel, {\it et al.,} Phys. Rev. Lett. {\bf 91}, 020402 (2003);
C. A. Regal, {\it et al.,} Nature {\bf 424}, 47 (2003); K.E. Strecker,
{\it et al.,} Phys. Rev. Lett. {\bf 91}, 080406 (2003); J. Cubizolles,
{\it et al.,} Phys. Rev. Lett. {\bf 91}, 240401 (2003); S. Jochim,
{\it et al.,} Phys. Rev. Lett. {\bf 91}, 240402 (2003); K. Dieckmann,
{\it et al.,} Phys. Rev. Lett. {\bf 89}, 203201 (2002); C.A. Regal,
{\it et al.,} Phys. Rev. Lett. {\bf 92}, 083201 (2004); M. Greiner,
{\it et al.,} Nature {\bf 426}, 537 (2003); M.W. Zwierlein, {\it et
al.,} Phys. Rev. Lett. {\bf 91}, 250401 (2003); S. Jochim, {\it et
al.,} Science {\bf 302}, 2101 (2003); M. Bartenstein, {\it et al.,}
Phys. Rev. Lett. {\bf 92}, 120401 (2004); T. Bourdel {\it et al.,}
cond-mat/0403091; C.A. Regal, {\it et al.,} Phys. Rev. Lett. {\bf 92},
040403 (2004); M.W. Zwierlein, {\it et al.,} Phys. Rev. Lett. {\bf
92}, 120403 (2004); C. Chin \etal Science {\bf 305},1128 (2004).


\bibitem{exp2} J. Kinast {\it et al.,} Phys. Rev. Lett. {\bf 92}, 150402
(2004); M. Bartenstein, {\it et al.,} Phys. Rev. Lett. {\bf 92},
203201(2004).

\bibitem{thomas} J. Kinast \etal cond-mat/0409283, see also \cite{levin}.

\bibitem{theory} E. Timmermans, {\it et al.,} Phys. Lett. A {\bf 285},
228 (2001); M. Holland, {\it et al.,} Phys. Rev. Lett. {\bf 87},
120406 (2001); Y. Ohashi and A. Griffin, Phys. Rev. Lett. {\bf 89},
130402 (2002) and references therein.

\bibitem{carlson} J. Carlson, \etal Phys. Rev. Lett. {\bf 91}, 050401
  (2003).

\bibitem{chang} S.-Y. Chang \etal Phys. Rev. A {\bf 70}, 043602 (2004).

\bibitem{giorgini} G. Astrakharchik \etal Phys. Rev. Lett. {\bf 93},
200404 (2004).

\bibitem{amoruso} M. Amoruso \etal Eur. Phys. J D {\bf 7}, 441 (1999);
G.M. Bruun and C.W. Clark, Phys. Rev. Lett. {\bf 83}, 5415 (1999);
M.A.  Baranov and D.S. Petrov, Phys. Rev. A {\bf 62}, 04601(R) (2000).

\bibitem{oscill} H.  Heiselberg, Phys. Rev. Lett. {\bf 93}, 250402
  (2004); C. Menotti \etal Phys. Rev. Lett. {\bf 89}, 250402 (2002);
  L. Vichi and S. Stringari, Phys. Rev. A {\bf 60}, 4734 (1999);
  M. Cozzini and S. Stringari, Phys. Rev. Lett. {\bf 91}, 070401
  (2003); H. Hu \etal Phys. Rev. Lett. {\bf 93}, 190403 (2004);
  M. Manini and L. Salasnich, cond-mat/0407039; Y.E. Kim and
  A.L. Zubarev, Phys. Lett. A {\bf 327}, 2004); Phys. Rev. A {\bf 70},
  033612 (2004); R. Combescot and X. Leynoras, Phys. Rev. Lett. {\bf
    93}, 138901 (2004); Europhys. Lett. {\bf 68}, 762 (2004).

\bibitem{stringari} S. Stringari, Europhys. Lett. {\bf 65}, 749
(2004).

\bibitem{abgfb05} A. Bulgac and G.F. Bertsch, Phys. Rev. Lett. {\bf
94}, 070401 (2005).


\bibitem{ho} T.-L. Ho, Phys. Rev. Lett. {\bf 92}, 090402 (2004).

\bibitem{heiselberg} H. Heiselberg, cond-mat/0409077.

\bibitem{levin} Q. Chen \etal cond-mat/0411090; this e-print has been
combined with the e-print cond-mat/0409283, \cite{thomas} and published
as joint paper, J. Kinast \etal Science {\bf 307}, 1296 (2005).

\bibitem{bdm} A. Bulgac \etal cond-mat/0505374.

\bibitem{fw} A.L. Fetter and J.D. Walecka, {\it Quantum Theory of 
Many-Particle Systems}, McGraw-Hill, San Francisco 1971.

\bibitem{chang0} J. Carlson \etal Phys. Rev. C {\bf 68}, 025802
(2003).

\bibitem{leggett} D.R. Eagles, Phys. Rev. {\bf 186}, 456 (1969);
A.J. Leggett, in {\it Modern Trends in the Theory of Condensed
Matter}, eds. A. Pekalski and R. Przystawa, Springer--Verlag, Berlin,
1980; J. Phys. (Paris) Colloq. {\bf 41}, C7--19 (1980); P. Nozi\`eres
and S. Schmitt--Rink, J. Low Temp. Phys. {\bf 59}, 195 (1985);
C.A.R. S\'a de Mello {\it et al.,} Phys. Rev. Lett. {\bf 71}, 3202
(1993); M. Randeria, in {\it Bose--Einstein Condensation},
eds. A. Griffin, D.W. Snoke and S. Stringari, Cambridge University
Press (1995), pp 355--392.

\bibitem{gorkov} L.P. Gorkov and T.K. Melik--Barkhudarov,
Sov. Phys. JETP {\bf 13}, 1018 (1961); H.  Heiselberg, {\it et al.,}
Phys. Rev. Lett. {\bf 85}, 2418 (2000).

\bibitem{alan} A. Griffin \etal Phys. Rev. Lett. {\bf 78}, 1838 (1997).

\bibitem{LL}E.M. Lishitz and L.P. Pitaevskii, {\it Statistical Physics}, 
part 1, Pergamon Press, Oxford, 1980.

\bibitem{cigar} A. Csordas and R. Graham, Phys. Rev. A {\bf 63}, 013606 (2001); 
{\it ibid} Phys. Rev. A {\bf 64}, 013619 (2001).

\bibitem{surf} C.J. Pethick and H. Smith, {\it Bose-Einstein Condensation 
in Dilute Gases}, Cambridge University Press, Cambridge, 2002;
L.P. Pitaevskii and S. Stringari, {\it Bose Einstein 
Condensation}, Oxford University Press, Oxford, 2003.


\end{thebibliography}
\end{document}